\documentclass[12pt]{article}

\usepackage{amsmath}
\usepackage{amsfonts}
\usepackage{amssymb}
\usepackage{amsthm}
\usepackage{setspace}
\usepackage{color}
\usepackage{subcaption}
\usepackage{caption}
\usepackage{enumerate}
\usepackage[hidelinks]{hyperref}
\usepackage{graphicx}
\usepackage[hidelinks]{hyperref} 
\usepackage[nottoc]{tocbibind}
\usepackage[utf8]{inputenc} 

\newcommand\ee{\end{equation}}
\newcommand\be{\begin{equation}}
\newcommand\eea{\end{eqnarray}}
\newcommand\bea{\begin{eqnarray}}

\newcommand\comment[1]{}
\newcommand\expect[1]{\left\langle #1 \right\rangle}

\comment{
\hypersetup{
colorlinks=true,
citecolor=DarkBlue,
linkcolor=DarkBlue,
urlcolor=DarkBlue,
}}

\def\O{\mathcal{O}}

\def\d{\partial}

\def\vphi{\varphi}

\def\ep{\epsilon}

\def\Mt{{\tilde M}}
\def\arccosh{{\rm arccosh}}
\def\arcsinh{{\rm arcsinh}}
\def\hyp{{}_2 F_1}
\def\ket{|{\rm ket}\rangle}
\def\bra{\langle {\rm bra}|}

\begin{document}

\begin{center}

\vglue 2cm
{\Large{\scshape  An Observer's Measure of De Sitter Entropy}}

\vskip 2 cm
{Mehrdad Mirbabayi}
\vskip 1 cm

{\em International Centre for Theoretical Physics, Trieste, Italy}

\vskip 2cm

\end{center}
\noindent {\bf Abstract:} {\small The two-point correlation function of a massive field $\expect{\chi(\tau)\chi(0)}$, measured along an observer's worldline in de Sitter (dS), decays exponentially as $\tau \to \infty$. Meanwhile, every dS observer is surrounded by a horizon and the holographic interpretation of the horizon entropy $S_{\rm dS}$ suggests that the correlation function should stop decaying, and start behaving erratically at late times. We find evidence for this expectation in Jackiw-Teitelboim gravity by finding a topologically nontrivial saddle, which is suppressed by $e^{-S_{\rm dS}}$, and which gives a constant contribution to $|\expect{\chi(\tau)\chi(0)}|^2$. This constant might have the interpretation of the late-time average of $|\expect{\chi(\tau)\chi(0)}|^2$ over all microscopic theories that have the same low-energy effective description.}

\vskip 1 cm
\newpage
\tableofcontents
\section{Introduction}
In a quantum system with a finite dimensional Hilbert space, correlation functions cannot decay forever. In the energy basis, every correlation function has an expansion in terms of purely oscillating phases, e.g.
\be\label{chi}
\expect{\chi(t)\chi(0)} = \sum_{ij} a_{ij} e^{-i(E_i-E_j) t},
\ee
where the information about whether $\expect{\cdot}$ means vacuum expectation value, thermal correlator, or something else is encoded in $a_{ij}$. At sufficiently short times, an overall decaying behavior can be mimicked by a conspiracy among different energy levels. But this collective behavior has to stop, if not earlier, by the time $t>1/\Delta E$, the typical spacing between the energy levels.

An observer in an inflating spacetime is surrounded by a horizon. In the limit of exact de Sitter, Gibbons and Hawking associated an entropy $S_{\rm dS}=A/4G_N$ to this horizon \cite{Gibbons}. Applying the holographic principle to the dS horizon suggests that there is a dual description in terms of a quantum system with a Hilbert space dimension $e^{S_{\rm dS}}$. If so, correlation functions measured along the observer worldline cannot decay indefinitely. However, effective field theory (EFT) predicts such an indefinite, exponential decay.

This contrast between the EFT prediction and the holographic principle, which is reminiscent of Hawking's information loss paradox \cite{Hawking}, has a counterpart for asymptotically AdS black holes. It was originally pointed out by Maldacena \cite{Maldacena_eternal}. The de Sitter version and the dramatic consequences of {\em macroscopic} Poincar\'e recurrences at exponentially long time scales ($t \sim e^{S_{\rm dS}}$ in units of dS curvature) were emphasized in \cite{Dyson1,Dyson2,Goheer}. Indeed, there is an $O(1)$ probability that in a landscape of vacua a given observer will tunnel out of any inflating vacuum before the macroscopic recurrence time \cite{CDL}.\footnote{In a quantum chaotic system there is another time-scale for {\em quantum} Poincar\'e recurrences after which correlation functions like \eqref{chi} could return arbitrarily close to their initial value. Typically this is doubly exponentially long in entropy. We thank Oliver Janssen for emphasizing this point.}

But even if inflation ends by reheating into an FLRW phase well before the Poincar\'e time, as long as it lasts longer than $t\sim S_{\rm dS}$, the FLRW observers will (naively) be able to collect much more information than $e^{S_{\rm dS}}$. This led the authors of \cite{Arkani} to introduce this exponentially shorter time-scale as the dS analog of the Page time for black hole evaporation \cite{Page_curve}. Remarkably, they showed that {\em slow-roll} models of inflation that satisfy the null energy condition and last for $t \gg S_{\rm dS}$ are inevitably eternal. This transition ensures the cosmological data set accessible to any succeeding FLRW observer is bounded by $e^{c S_{\rm dS}}$ with $c=\O(1)$. Subsequent works \cite{Creminelli,Dubovsky1,Dubovsky2} have sharpened the bound and obtained an intriguingly universal threshold $c=1/2$.

We are left with several open problems: What follows the exponentially decaying phase of the correlation functions? Holography suggests there has to be a transition before they become exponentially small in $S_{\rm dS}$, namely at $t\sim S_{\rm dS}$. But will there be another coherent phase before reaching the Poincar\'e time $t\sim e^{S_{\rm dS}}$? What is the significance of the sharp threshold $c=1/2$? Is there a similar bound on the data collected by FLRW observers succeeding {\em false-vacuum} eternal inflation (a scenario that was left unconstrained by \cite{Arkani})?

Our goal in this work is to look for a sign of the break-down of EFT within the inflating universe. If successful, this might shed some light on the above questions, and the holographic picture that underlies them. However, at first sight, we are embarking on an ill-motivated effort since, on the one hand, formulating sharp questions often fails when and where dynamics of gravity cannot be treated as a small perturbation, and, on the other, the supposed break-down of EFT is exactly because we are dealing with non-perturbative gravitational dynamics.

What encourages us to pursue this is the recent progress in defining observables in quantum gravity in relation to a dynamical observer \cite{Chandra,Witten}. The relational nature of observables in quantum gravity is of course a well-known fact. As Coleman puts it \cite{Coleman}:

\vskip 0.5 cm

\noindent {\em When gravity becomes part of the dynamics [\dots] instead of asking, ``At four o'clock, what is the probability distribution of electron position?'', we can ask ``If we project the wave function of the universe on to the subspace in which the hands of the clock are pointing to four, what is the probability distribution of electron position?''}

\vskip 0.5cm
\noindent What is new is the striking simplicity of the formalism with which \cite{Chandra,Witten} implement this idea and the evidence they provide for the holographic interpretation of $S_{\rm dS}$. Although they work in the $G_N\to 0$ limit, and the evidence they provide is really for the holographic interpretation of changes in the horizon entropy under $\O(G_N)$ changes in the horizon area, we see no immediate obstruction in using their formalism to compute correlation functions measured by an observer even at finite $S_{\rm dS}$. So we will pursue this and ask if the EFT prediction for the correlation functions breaks down at large proper time separation.
\section{Dynamical observer with a clock}\label{sec:obs}
Including a dynamical worldline in the gravitational path integral allows to partially localize fields along the worldline, but to fully localize them we need an additional clock degree of freedom. The proposal of \cite{Chandra,Witten} is to supplement the observer worldline coordinates $X^\mu$ with a pair of conjugate variables $(p,q)$, with the action
\be
I_{\rm obs}=  \int d\tau \left[p \frac{dq}{d\tau} -(M+q) \sqrt{-g_{\tau\tau}}\right],
\ee
and the constraint that the clock Hamiltonian $q$ is bounded below, ${\rm min}(q)>-M$. Denoting the matter fields collectively by $\chi$, the gravitational path integral now looks
\be\label{Z}
Z = \int Dg_{\mu\nu} D\chi DX^\mu Dp Dq e^{iI[g_{\mu\nu},\chi]+iI_{\rm obs}[g_{\mu\nu},X^\mu,p,q]}.
\ee
Classically, the ``clock'' variable $p$ coincides with minus the proper time of the observer. Diff-invariant correlation functions of a bulk scalar field $\chi$ at proper time $-p_1$ can be defined by inserting
\be
\int d\tau   \frac{dp}{d\tau} \delta(p(\tau)-p_1) \ \chi(X^\mu(\tau))
\ee
in the path integral. (To measure spinning operators, one needs to also introduce a dynamical orthonormal frame along the worldline.) In practice, reparametrizations of $\tau$ can be fixed by setting $dp/d\tau = -1$, thereby eliminating $p$ in favor of $\tau$ (or vice versa). Then we can simply talk about $\expect{\chi(X^\mu(\tau_1))\chi(X^\mu(\tau_2))}$ as a well-defined correlation function along the worldline. 

Perhaps this last step makes the introduction of a dynamical clock look superfluous. Of course, it is conceptually important because it allows the rest mass of the observer $M+q$ to change (and total $T_{\mu\nu}$ remain locally conserved) when operators are inserted along the worldline. This is a necessary requirement when gravity is dynamical, in contrast to the non-gravitational case where energy and momentum can by injected and extracted by local insertions. In practice, though, we are usually interested in cases where the effect is negligible, like when we neglect the effect of detecting CMB photons on the orbit of the Earth. Below we will see an example that this natural expectation breaks down at late times.

Having introduced an {\em algebra} of well-defined observables with respect to a dS observer, the authors of \cite{Chandra,Witten} succeeded in showing an interesting property of this algebra. Namely that, in the limit $G_N\to 0$, it admits a notion of entropy up to a (possibly infinite) additive constant. And that this entropy is maximized in empty dS. This indeed agrees with assigning an entropy to the horizon, i.e. taking the total entropy in the static patch (the spacetime region that is causally connected to the observer) to be
\be
S_{\rm tot} = \frac{A}{4G_N} + S_{\rm matter}.
\ee
Empty dS is believed to maximize $S_{\rm tot}$ because the positive entropy of matter excitations is overcompensated by the shrinking of the horizon area \cite{Maeda,Bousso}. 

Our interest is in finite $S_{\rm dS}$. We will nonetheless continue to use the same continuous description of the clock. This might sound inconsistent with the holographic principle since a realistic clock inside the static patch must have a Hilbert space dimension much less than $e^{S_{\rm dS}}$. However, this is the same inconsistency that the whole semi-classical description \eqref{Z}, with continuous fields on a smooth geometry, suffers from. Because we are going to use the semi-classical gravity to diagnose a transition at late times, using a continuous model of the clock actually seems appropriate. Of course, the entire program might fail.
\section{dS solution in JT gravity}
We will work with a $2d$ toy model called Jackiw-Teitelboim (JT) gravity \cite{Jackiw,Teitelboim}, hoping that it captures the salient features of the actual higher dimensional problem. It is actually a dilaton-gravity model, and hence not a purely topological theory. The model was studied in detail in \cite{Maldacena_dS}. This section is a short review of the relevant results. The gravitational action is given by
\be
I_{g}[\phi,g_{\mu\nu}] = \frac{S_{\rm dS}}{8\pi} \left[\int d^2 x \sqrt{-g} R - 2 \int_{\rm Bdy} K\right]
+\int d^2 x\sqrt{-g} \phi (R-2)- 2 \phi_b \int_{\rm Bdy} K.
\ee
The Einstein-Hilbert action plus its boundary term, inside the square brackets, is topological. It gives $4\pi$ times the Euler characteristic of the manifold. The path integral over $\phi$ fixes $R=2$, and the one on the metric ensures $\phi = \phi_b$ at the boundary. This action admits ``global'' dS solutions of the form
\be\label{dS2}
ds^2 = -dt^2 + \cosh^2 t \ d\vphi^2, \qquad \phi = \phi_r \sinh t,
\ee
where $\phi_r$ is an integration constant, which determines the length of the late time boundary $\ell$. In the limit $\phi_b,\ell \gg 1$, 
\be\label{phir}
\phi_r \approx \frac{2\pi \phi_b}{\ell}.
\ee
\begin{figure}[t]
  \centering
  \includegraphics[scale =1.]{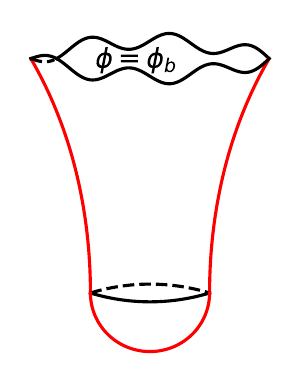} 
  \caption{\small{The Hartle-Hawking prescription for the wavefunction. With this prescription, any observer maximally extends into two (the red curve). }}
  \label{fig:hh}
\end{figure}
We think of the late time boundary $\phi = \phi_b$ as an analog of the reheating surface. One can imagine turning on interactions and changing the description afterward. Hence, larger $\ell$ means longer inflation, with the total number of inflationary e-folds given by $N_e \approx \log \frac{\ell}{2\pi}$.

The Hartle-Hawking state \cite{HH} is defined by performing the gravitational path integral on complex geometries with no other boundary than the one at $\phi = \phi_b$ (see figure \ref{fig:hh}). The saddle-point of this integral is given by the above solution, but with $t$ taken as a complex variable that moves along the following contour
\be\label{hh_contour}
\includegraphics[scale =0.9]{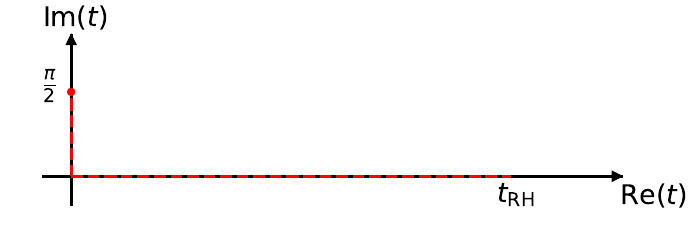} 
\ee
The purely imaginary part $t= it_E$ describes a hemisphere
\be\label{S2}
ds^2 = dt_E^2 + \cos^2 t_E d\vphi^2,\qquad 0\leq t_E\leq \frac{\pi}{2},
\ee 
that smoothly connects to \eqref{dS2} at $t=0$. We have normalized the action such that this saddle-point approximation gives $e^{S_{\rm dS}}$ for the norm-squared of the Hartle-Hawking state, as in higher dimensional cases. 

The matter fields are minimally coupled to the $2d$ metric, and so they are indirectly coupled to and back-react on the dilaton. We further simplify our computation by considering heavy fields $m\gg 1$, whose correlation functions can be approximated by point-particle worldlines. In this limit, matter fields cause a discontinuity in the derivative of dilaton normal to the worldline 
\be\label{junc0}
\phi_+ = \phi_-,\qquad \xi^\mu (\d_\mu\phi_+ - \d_\mu \phi_-) =\frac{1}{2} m,
\ee
where $\phi_\pm$ are the values of the dilaton field on the two sides, and $\xi^\mu$ is the normal pointing from the $+$ side to the $-$ side. 

In addition to the matter fields, we are including an observer with rest mass $M+q\gg 1$, who will similarly back-reacts on $\phi$. We are interested in a situation where, at least at short times, this back-reaction is small, so the observer can be considered a test particle moving along a geodesic of \eqref{dS2}, e.g.
\be
\vphi = 0.
\ee
The change in the proper time $\tau$ of this observer (and any other with $\vphi=$ constant) coincides with the change in $t$ coordinate. Note that in the Hartle-Hawking state we always have pairs of observers whose worldlines connect in the Euclidean regime as in figure \ref{fig:hh}.

As discussed in \cite{Maldacena_dS}, the above $2d$ model can be obtained from dimensional reduction of near-extremal geometries in higher dimensions. Such geometries normally develop a long throat such as dS$_2 \times S^2$, with $\phi$ characterizing the variation in the area of $S^2$ normalized by $G_N$. The number of e-folds in the $2d$ model would then have to do with how close the original solution is to extremality, and $\phi_b$ corresponds to an $\O(1)$ relative variation in the area. Therefore, with our normalization $\phi_b \sim S_{\rm dS}$ and
\be
\phi_r \sim S_{\rm dS}e^{- N_e}.
\ee
Unfortunately, this does not cover the regime we are interested in, i.e. $N_e \approx t_{\rm RH}\sim S_{\rm dS}$. A necessary condition to have small back-reaction, given \eqref{junc0} and $m, M\gg 1$, is $\phi_r\gg 1$, which implies $N_e <\log(S_{\rm dS})\ll S_{\rm dS}$. The time scale $\log S$ is known as the scrambling time in black hole physics \cite{Maldacena_chaos}, but it does not seem to be a meaningful limit for inflation, at least not in $2d$. We will therefore consider $2d$ parameters that significantly deviate from what dimensional reduction suggests, namely $\phi_b \gg S_{\rm dS}$, in order to have a more faithful analog of inflation in higher $d$. Can this regime be realized in a more elaborate compactification scenario is an interesting question.
\section{Matter correlator without back-reaction}
In de Sitter, we are usually interested in the in-in correlation functions. They are obtained by evolving forward in time from an initial state (or from a complex manifold with no boundary in the past), inserting operators as they are written from right to left, appropriately evolving forward or backward in time if these operators have different time arguments, and finally returning to the initial state.

The easiest case (and the only one we consider) is the 2-point correlation function of a scalar field 
\be
G_\chi(\tau) = \expect{\chi(X)\chi(Y)},
\ee
where $\tau$ is the geodesic time from $Y$ to $X$, with ${\rm Im}(\tau)<0$. We don't lose much generality by assuming $X$ is to the future of $Y$; the opposite case is related by complex conjugation. So we can think of preparing a $\ket$ by forward evolution and insertion of $\chi(Y)$ and $\chi(X)$, and then taking the overlap of the resulting state with a $\bra$ that was prepared with no operator insertions. This can be depicted as follows
\be
\includegraphics[scale =0.8]{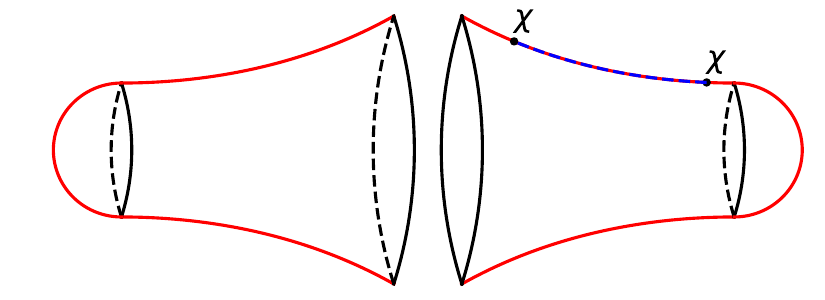} 
\ee
We will give the exact expression for $G_\chi(\tau)$ in the appendix. For our discussion, all we need to know is that for heavy fields, at large separation ($\tau = t-i\ep$ with $t\gg 1$)
\be\label{2pf}
G_\chi(t-i\ep) \sim e^{-\frac{1}{2}t -i\nu t},\qquad \nu \equiv \sqrt{m^2 - \frac{1}{4}},
\ee
and that this is describing a point particle moving along a geodesic. The factor $e^{-t/2}$ gives the spread of the wavefunction due to the expansion of the universe, and it arises from the 1-loop determinant around the geodesic saddle. The connection between dS correlators and geodesics and the possibly important role they might play in dS holography has been recently emphasized in \cite{Chapman,Aalsma}.

It is useful to analytically continue the two-point correlator to complex geodesic separation $\tau=t -i\pi$ 
\be
\label{2pf_sl}
G_\chi(t -i\pi) \sim e^{-\frac{1}{2}t-\pi \nu} \cos\left(\nu t -\frac{\pi}{4}\right).
\ee
This describes the space-like configuration where we have two insertions along the two observer worldlines connected via the Hartle-Hawking prescription:
\be\label{braket2}
\includegraphics[scale =0.8]{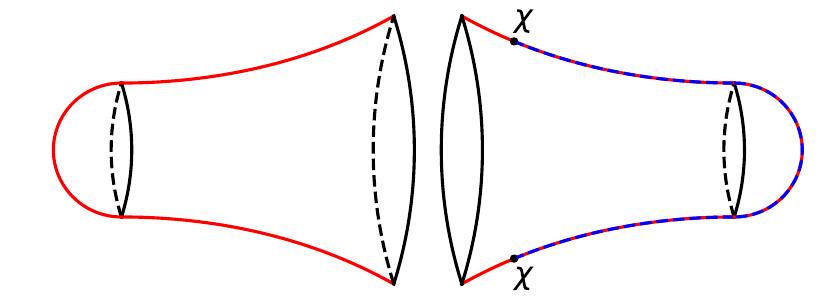} 
\ee
Cosine arises from the fact that there are two geodesics, one fully on the $\ket$ side, as in the above figure, and the other connecting the two $\chi$'s by extending in the $\bra$ side. Both of these have a Euclidean section of length $\pi$, and hence contribute with equal weight. 

Both \eqref{2pf} and \eqref{2pf_sl} decay exponentially with $t$. If they continue to do so well past $t \sim S_{\rm dS}$ they cannot be described by a holographic dual with $e^{S_{\rm dS}}$ degrees of freedom. The alternative is that the EFT computation is breaking down. Sometimes, when formulating the right question, one EFT computation shows the break-down of the other. The prototypical example is the recent success in obtaining the Page curve for the entropy of a thermal bath coupled to a black hole \cite{Almheiri,Penington}. While EFT predicts thermal Hawking radiation and hence an ever growing entropy, direct computation of entropy using the gravitational path integral shows a transition at the Page time $t\sim S_{\rm BH}$ due to a shift of the dominance of the saddle points. The new saddle is suppressed by $e^{-S_{\rm BH}}$. It dominates because the contribution of the old saddle decays exponentially with time.

There is some evidence that quantities that average over microscopic details (such as the total entropy) might be reliably computed using the EFT, even when it fails because other observables become sensitive to those details. The gravitational description might give an answer for $G_\chi(\tau)$ that is incompatible with holography, but it might correctly show the sign of a transition in $|G_\chi(\tau)|^2$. The idea being that when the former starts showing an erratic behavior, sensitive to the detailed structure of the microstates, the average behavior of the latter is dominated by a new saddle. This very much resembles the computation in \cite{Stanford}. 
\section{Other manifolds and other geodesics}
Given that the observer moves along a geodesic (away from the operator insertions) and matter correlation functions can be approximated using geodesic distances, our objective in this section will be to find a geometry, and on that geometry appropriate geodesics that could give a non-decaying contribution to $|G_\chi(\tau)|^2$. For now we won't worry about back-reaction -- namely, whether or not dilaton equations are satisfied -- but return to it in the next section.
\subsection{The disk (Hartle-Hawking)}
A nice trick to look for new geometries is analytic continuation to AdS \cite{Maldacena_NG,Maldacena_dS}. Making the following substitution in \eqref{dS2}
\be\label{anal}
t = i\frac{\pi}{2} + r,\quad \phi_r = -i A,
\ee
results in 
\be\label{ads2}
ds^2 = - (dr^2 + \sinh^2 r d\vphi^2),\qquad \phi = A \cosh r. 
\ee
Up to an overall sign, this is the metric of the Euclidean AdS$_2$. It is often visualized by the conformal map $\hat r = \tanh\frac{r}{2}$ to the unit disk. The dilaton solution can be obtained from the dimensional reduction of the near-horizon throat of a near-extremal charged black hole \cite{Maldacena_AdS}. $r=0$ is the horizon. As in the dS case the geometry is cut off when $\phi =\phi_b$, which corresponds to
\be
r_{\rm UV} \equiv \arccosh \frac{\phi_b}{A}.
\ee
Under this analytic continuation, the observer worldline maps to a geodesic that runs right through the horizon $r=0$, e.g. $\{\vphi =0\}\cup\{\vphi = \pi\}$, and connects two boundary points that are separated by a distance $2 r_{\rm UV}$. Under \eqref{anal}, we obtain the expected result
\be\label{Deltau}
\Delta \tau = -i\pi + 2 t_{\rm RH},
\ee
where on the dS side, the ``reheating'' time $t_{\rm RH}$ is defined as $\phi_r\sinh t_{\rm RH} = \phi_b$. 

So far we haven't found anything new. In fact, we can think of the disk geometry as a deformation of the contour \eqref{hh_contour} that defines the Hartle-Hawking saddle. More general geodesics, still on AdS disk, have a minimum impact parameter $d$:
\be
\includegraphics[scale =1.]{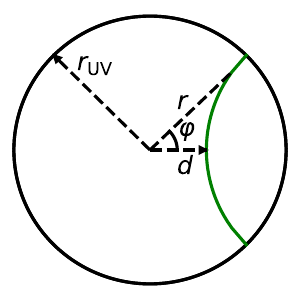} 
\ee
Measuring $\vphi$ from the direction of the minimum approach, they follow
\be\label{boosted}
\cos \vphi = \frac{\tanh d}{\tanh r},
\ee
Upon the analytic continuation \eqref{anal}, this describes a pair of time-like geodesics that are boosted with respect to the constant $\phi$ slices
\be
\vphi_\pm =\pm \arccos( \tanh d \ \tanh t).
\ee
They start at angular separation $\pi$ and asymptotically reach $\Delta \vphi_{\rm min} = 2\arccos(\tanh d)$. They are related to the stationary observers in figure \ref{fig:hh} by a dS isometry, and hence can be connected by a purely Euclidean piece of length $\pi$. Conversely, any pair of superhorizon points are connected by one such complex geodesic \cite{Chapman,Aalsma}. 
\subsection{The double trumpet (wormhole)}
In the JT path integral, higher genus geometries are suppressed by powers of $e^{-S_{\rm dS}}$. However, to get an answer for $|G_\chi(\tau)|^2$ that differs from just squaring the answer for $G_\chi(\tau)$, we have to consider geometries that connect the two factors. So we need to pay this price. The least suppression comes from joining the $\ket$ of $G_\chi(\tau)$ to the $\bra$ of $G_\chi(\tau)^*$, keeping the remaining $\ket$ and $\bra$ the same as the Hartle-Hawking state. Geometries that connect the $\ket$ and $\bra$ have been proposed previously by Page \cite{Page_rho}. Two recent works that discuss them in the context of JT gravity are \cite{Penington} and \cite{Chen}. The focus of \cite{Chen} is on geometries that connect the $\bra$ and $\ket$ of the same correlator, dubbed ``bra-ket wormholes''. Our solution is similar to one of theirs. 

A geometry that connects $\ket$ and $\bra$ has two expanding sides. On the Euclidean AdS$_2$ side, this is known as the double trumpet \cite{SSS2}. Its metric (after multiplying by a negative sign to satisfy $R=2$) is given by
\be\label{dt}
ds^2 =-(d\rho^2 + \cosh^2 \rho \ dy^2),\qquad -\infty<\rho<\infty,\qquad y\sim y + b,
\ee
where the periodicity $b$ is, for now, a free parameter. This geometry can be obtained from the disk by identifying two geodesics with closest distance $b$:
\be
\includegraphics[scale =1.]{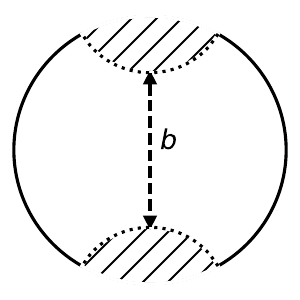} 
\ee
There is no vacuum solution of the dilaton equation that is periodic under $y\to y+b$ and increases in both positive and negative $\rho$ directions to satisfy $\phi = \phi_b$ at the UV boundaries. We will see that there is a solution, with $b$ stabilized, after the back-reaction of the observer and the $\chi$ field is taken into account. Since the metric remains the same, it is illustrative to see how correlating fields through this wormhole leads to a non-decaying late-time answer. 

First, to analytically continue to dS, we set
\be\label{analrho}
\rho = i \left(n-\frac{1}{2}\right)\pi + t,\qquad n\in {\mathbb Z},
\ee
in the double trumpet \eqref{dt}, and find 
\be\label{Milne}
ds^2 = -dt^2 + \sinh^2 t \ dy^2, \qquad y\sim y+b.
\ee
\begin{figure}[t]
  \centering
  \includegraphics[scale =0.8]{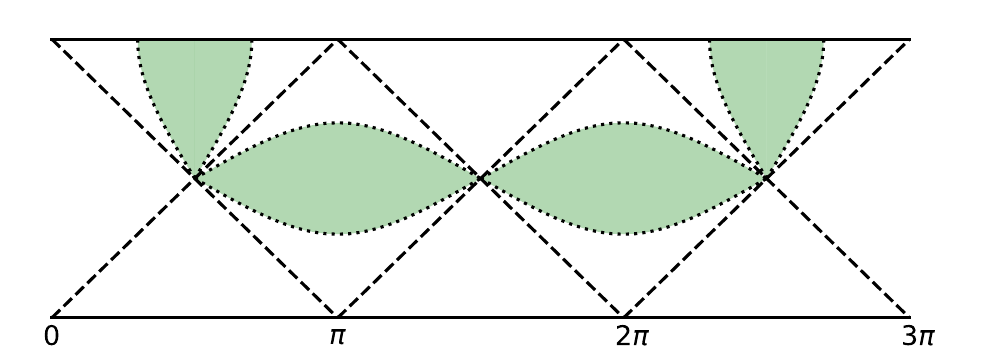} 
  \caption{\small{The Lorentzian parts of the geometry that corresponds to the contour \eqref{rho_contour}, embedded in the Penrose diagram of the global cover of dS$_2$. The dotted lines within each wedge are identified. The conical singularities are avoided by contour deformations at $0$,$i\pi$ and $2i\pi$. }}
  \label{fig:cones}
\end{figure}
This is a Lorentzian cone, describing a compactification of the future lightcone or the past lightcone of a point on global dS$_2$ geometry. Without the compactification it would be just the Milne slicing of dS$_2$. On dS$_2$ geometry \eqref{dS2} the periodicty $\vphi\sim \vphi+2\pi$ implies $n \sim n+2$ in \eqref{analrho}, and there are four cones to the future and past of two antipodal points. If $\vphi$ is decompactified then different $n$ become inequivalent. As in the Hartle-Hawking case where we choose the contour \eqref{hh_contour} to define a complex manifold in the space of complex metrics, here we choose the following contour for $t$, and obtain the geometry shown in figure \ref{fig:cones}.\footnote{We thank Yiming Chen for pointing out that on the contour considered in the previous version of this paper, dilaton would not satisfy the right boundary condition. The new contour is the same as the $2\pi$ contour in \cite{Chen}, but the two possible ways to aviod the singularity at $t = i\pi$ do not seem to change our result.} 
\be\label{rho_contour}
\includegraphics[scale =1.]{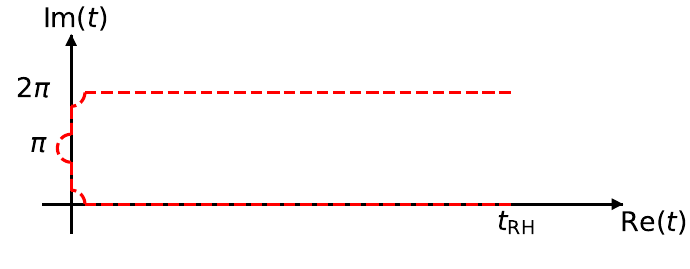} 
\ee

Next consider lines of $y=$ constant, which obviously are geodesics. Since they stretch between the two asymptotic regions, they are our candidates for giving the correlators of $\chi$ between $\ket$ and $\bra$. When stretched between points $t$ and $t+2i\pi$ on \eqref{rho_contour}, their length is purely spacelike, equal to $2i\pi$. As a result, there won't be any exponential suppression of the $G_\chi(\tau)$ as $t \to \infty$. 

\begin{figure}[t]
  \centering
  \includegraphics[scale =1.]{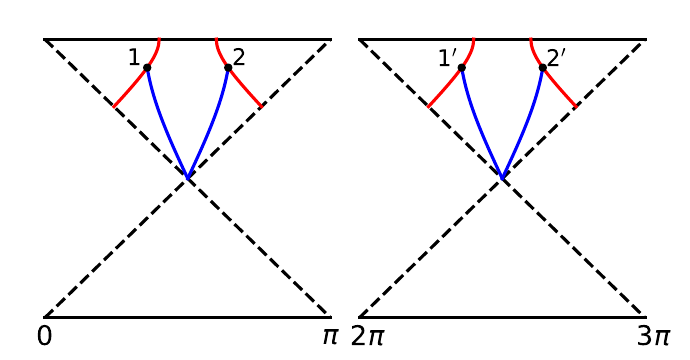} 
  \caption{\small{Analytic continuation of the geodesics in \eqref{trumpet_geod}, on the global cover of the expanding cones of figure \ref{fig:cones}. $y=$ constant geodesics (blue) connect points on the two sides: $1\to 1'$ and $2\to 2'$. Their length is Euclidean, $2i\pi$. This is consistent with the expectation that the accumulated phase during the Lorentzian propagation should cancel between the $\ket$ and the $\bra$. Each red curve in \eqref{trumpet_geod} gives the pair of observers either on the $\ket$ or $\bra$. }}
  \label{fig:penrose_geod}
\end{figure}

Of course, we need to anchor $\chi$ insertions on the observer worldline. On the double trumpet this corresponds to a configuration like
\be\label{trumpet_geod}
\includegraphics[scale =1.]{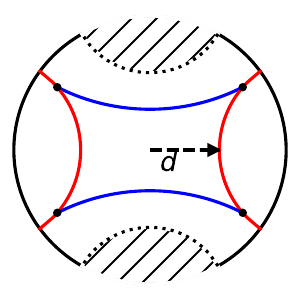} 
\ee
such that the observer worldline (red) stretches between two points on the same asymptotic boundary. Such geodesics have either $\rho>0$ or $\rho<0$ and are characterized by ${\rm min}(|\rho|) = d$, and a trivial translation along $y$. Setting that to zero, we have 
\be
\cosh y = \frac{\tanh \rho}{\tanh d},
\ee
and its reflection for the $\rho<0$ curve. The geodesic distance between two symmetric points $(\rho,\pm y)$ is $\Delta \tau = 2 \arccosh \frac{\sinh \rho}{\sinh d}$. Each of these geodesics analytically continues to a pair of boosted observers. In the global embedding, this is the same pair described below \eqref{boosted}, and shown in figure \ref{fig:penrose_geod}. In the coordinates of \eqref{Milne}, they asymptote to 
\be
y(t\to \infty) =\pm \arccosh \frac{1}{\tanh d},
\ee
with geodesic separation
\be
\Delta \tau = - i \pi + 2 \arcsinh \frac{\cosh t_{\rm RH}}{\sinh \rho_m}\sim -i\pi + 2 t_{\rm RH}.
\ee
Therefore, we can imagine far separated insertions of $\chi$ along the observer worldlines as required for computing $|G_\chi(\tau)|^2$ at large ${\rm Re}(\tau)$. On the wormhole geometry, this will not decay because the distance between the $\chi$ insertions on the $\ket$ and the $\bra$ approaches a finite limit. 

\section{Matter correlator with back-reaction}
In this section, we take into account the back-reaction of the observer with mass $M$ and the field $\chi$ with mass $m$ on the dilaton. We work in the regime where this is a small effect on the Hartle-Hawking (or the disk) saddle. Nevertheless, back-reaction is what ensures the existence of a dilaton solution on the wormhole (the double trumpet). 
\subsection{Back-reaction on the disk}
We will solve the dilaton equation \eqref{junc0} via analytic continuation to negative AdS. Away from the matter sources, the solution for the dilaton is always as in \eqref{ads2}, i.e. $A\cosh r$. The effect of the massive worldlines is to change $A$ on the two sides, and the position of the ``horizon'', namely the point from which the geodesic distance $r$ is measured. Let us denote by $A_+$ and $A_-$ the amplitudes on the two sides of a worldline, and by $d_+$ and $d_{-}$ the shortest distance between the two horizons and the curve:
\be\label{back_disk}
\includegraphics[scale =1.]{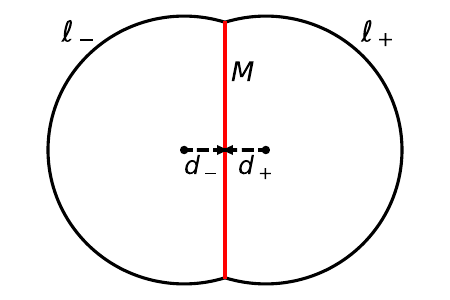} 
\ee
The continuity condition, $\phi_+ = \phi_-$, along the worldline implies that the same point has the shortest distance to both horizons, and therefore \eqref{junc0} reduces to 
\be\label{junc}\begin{split}
A_+ \cosh d_+ - A_- \cosh d_- & =0 ,\\[10 pt]
A_+ \sinh d_+ + A_- \sinh d_- &= -\frac{i}{2}M,
\end{split}\ee
where the factor of $i$ comes from the normalization of the normal to the worldline $\xi^\mu$; it is a spacelike vector on a negative signature metric \eqref{ads2}. Here we assumed the two horizons are on opposite sides of the curve, as shown in \eqref{back_disk}. A negative $d_-$ implies both are on the same side. Back-reaction affects the lengths of the boundary curves $\phi = \phi_b$ on the two sides of the geodesic, i.e. the lengths of the two pieces of the reheating surface separated by the end points of the maximal extension of the observer worldline. Using \eqref{boosted} for the angular size of the geodesic, we find
\be\label{ell}
\begin{split}
\ell_+ &= 2(\pi - \arccos(\tanh d_+)) \frac{\phi_b}{-iA_+},\\[10pt]
\ell_- &= 2 (\pi -\arccos(\tanh d_-)) \frac{\phi_b}{-iA_-}.\end{split}
\ee
The factors of $-i$ are because the boundary conditions are imposed in the asymptotic future of dS, namely when the metric looks like $ds^2 = -dt^2 + e^{2t} dx^2$. Asymptotic behavior in negative AdS is $ds^2 = -dr^2 - e^{2r} dx^2$, $\phi = A e^r$, so we have to send $r\to -i\frac{\pi}{2} +t$.

To summarize, given $\ell_+$ and $\ell_-$ and $M$, the dilaton solution is fully fixed by (\ref{junc},\ref{ell}). In particular, for an unboosted observer, $\ell_+ = \ell_- = \ell/2$, and assuming small back-reaction, $M\ll -iA_\pm$, we find
\be\label{phir2}
-iA_\pm \approx \frac{2\pi \phi_b}{\ell} = \phi_r,
\ee
where we used the definition of $\phi_r$ in \eqref{phir}.

The $\chi$ insertions along the worldline have no effect on the dilaton solution on this saddle. In the worldline approximation, we are adding a particle of mass $m$ moving along the same geodesic as the observer. By energy-momentum conservation, in the interval between the two $\chi$ insertions the observer rest mass reduces to $M-m$, and hence the same mass $M$ appears in \eqref{junc}. Somewhat intuitively, in order to be able to measure the correlator, the observer has to start in an excited state. In the language of section \ref{sec:obs}, we are taking ${\rm min} (q) <-m$. 
\subsection{Back-reaction on the double trumpet}
On the double trumpet geometry, $\chi$ fields can be contracted between the two observer worldlines. This has a nontrivial effect on the dilaton. We will find a solution that is symmetric between the $\ket$ and $\bra$, and with symmetric insertions of $\chi$ along the observer worldline, as in figure \ref{fig:penrose_geod}. This is the simplest configuration, appropriate for computing $|G_\chi(-i\pi + t)|^2$. We will comment on the case with asymmetric $\chi$ insertions in the end.

\begin{figure}[t]
  \centering
  \includegraphics[scale =0.9]{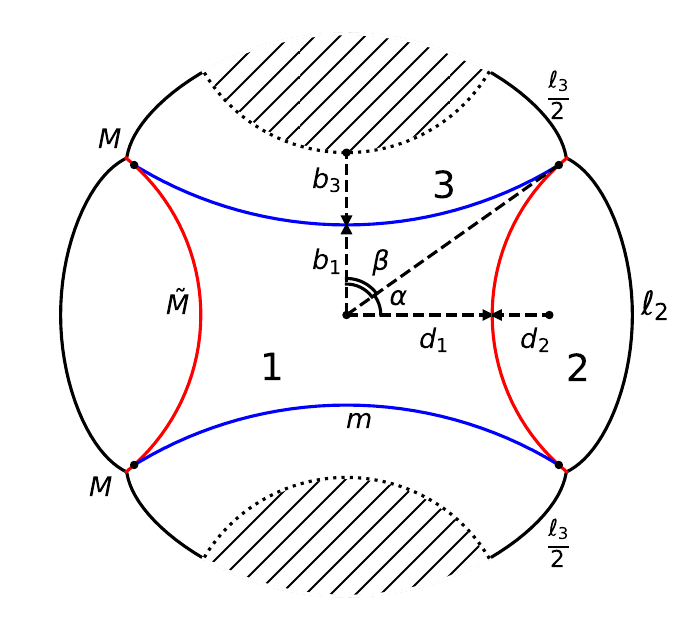} 
  \caption{\small{Double trumpet geometry, including the matter back-reaction on the dilaton. }}
  \label{fig:dilaton_tr}
\end{figure}

First, we can fix the dilaton amplitudes $B_{1,2,3} \equiv -i A_{1,2,3}$ and the length scales $d_1,d_2$, $b_1,b_3$ shown in figure \ref{fig:dilaton_tr} in terms of the boundary lengths $\ell_2,\ell_3$, $M$ and $m$. Then we identify $\ell_2, \ell_3$ with $\ell_\pm$ in \eqref{back_disk} since we are computing the overlap between the $\bra$ and $\ket$ that are connected via a wormhole, and the other $\bra$ and $\ket$ that are given by two disks.

By symmetry, $b_1$ and $d_1$ are perpendicular to each other. Consider the geodesic connecting the horizon $1$ to a $\chi$ insertion and denote by $\alpha$ and $\beta$ the angles it makes with $d_1$ and $b_1$. If the $\chi$ insertions are sent to infinity (which is the limit of interest), it follows from \eqref{boosted} that $\alpha$ and $\beta = \frac{\pi}{2}-\alpha$ satisfy
\be
\cos \alpha \approx \tanh d_1,\qquad \cos\beta \approx \tanh b_1,
\ee
implying $\sinh d_1 = 1/\sinh b_1$. In the same limit, the geodesic of the $\chi$ particle gets emitted almost collinear with the observer worldline. Therefore, the observer mass after the emission is $\Mt \simeq M - m$. Using these and the dilaton equations, we find
\be
\tanh d_2 =\frac{\Mt}{2 B_1}\tanh b_1- \frac{1}{\cosh b_1}   ,\qquad \tanh b_3 = \frac{m}{2 B_1 \cosh b_1}-\tanh b_1,
\ee
and
\be
B_2 =\sqrt{B_1^2 - (\Mt/2)^2+ \Mt  B_1/\sinh b_1},\qquad B_3  =  \sqrt{B_1^2 -(m/2)^2 +m B_1 \sinh b_1}.
\ee
Now we can impose the boundary conditions on the length of the two segments of the reheating surface:
\be\label{trumpet}\begin{split}
\frac{B_1\ell_{2}}{\phi_b } &\approx \frac{2}{\sqrt{1-\frac{(M-m)^2}{4B_1^2} +  \frac{M-m}{B_1 \sinh b_1}}}
\left[\pi - \arccos\left(\frac{M-m}{2B_1} \tanh b_1-\frac{1}{\cosh b_1}\right)\right],\\[10 pt]
\frac{B_1\ell_{3}}{\phi_b } &\approx \frac{1}{\sqrt{1-\frac{m^2}{4B_1^2} +  \frac{m}{B_1} \sinh b_1}}
\left[\pi - 2 \arccos\left(\frac{m}{2B_1 \cosh b_1}- \tanh b_1 \right)\right].
\end{split}\ee
Given $\ell_2$ and $\ell_3$ this can be solved for $B_1$ and $b_1$. In the limit of small back-reaction, $M/B_1<1$ and $m\ll M$, the equations simplify. In particular, considering an unboosted observer on the original disk saddle, which implies $\ell_2 = \ell_3=\ell/2$, we find
\be
b= 2(b_1 + b_3) \approx \sqrt{\frac{2\pi m}{\phi_r}}\ll 1,\qquad B_1 \approx \sqrt{\frac{\phi_r m}{2\pi}},
\ee
where $\phi_r = 2\pi \phi_b/\ell$.
\subsection{Large back-reaction on the observer}
A modification of this problem is interesting to consider: the observer can measure the time at which $\phi =\phi_b$, when reheating happens. Hence, we can impose a constraint that the total length of the observer worldline, maximally extended, is fixed to $T$. In the notation of section \ref{sec:obs}, before gauge-fixing $p= -\tau$, this is
\be
-\int d\tau \dot p = T.
\ee
Imposing this condition allows a change in the conjugate variable $q$ and hence different masses of the observer on the disk versus the double trumpet.

As always we first study the disk. When the original mass of the observer $M$ is small enough that its back-reaction on the disk geometry is negligible, and assuming $\ell_+ = \ell_-=\ell/2$, the two horizons are located on opposite sides and close to the observer worldline, i.e. $d_\pm\ll 1$. Substituting in the identity
\be\label{Pythagoras}
\cosh r_{\rm UV} = \cosh d_\pm \cosh \frac{T}{2},
\ee
we find
\be
T = 2 \arccosh \left(- i\frac{\phi_b}{\phi_r}\right) \approx -i\pi + 2 t_{\rm RH}.
\ee

On the wormhole side, we solve \eqref{trumpet} but with $M\to M_w$, which could differ from $M$, and still assuming $m\ll M_w$. For the $\ell_2,\ell_3$ and the total observer time to agree with those on the disk side, the angular size of the observer worldline as measured from horizon 2 in figure \ref{fig:dilaton_tr} has to be $\approx \pi$. Using \eqref{boosted}
\be
\arccos(\tanh d_2)= \arccos\left(\frac{M_w}{2B_1}\tanh b_1 -\frac{1}{\cosh b_1}\right) \approx \frac{\pi}{2},
\ee
which implies
\be
M_w \approx \frac{2B_1}{\sinh b_1}.
\ee
We still assume $m\ll B_1$, which in order to have $b_3>0$ implies $\sinh b_1 < m/(2B_1)\ll 1$. Then the following approximate relations hold
\be
\ell_2 \approx \frac{\pi \phi_b b_1}{B_1},\qquad 
\ell_3 \approx \frac{2\phi_b}{B_1}\left(\frac{m}{2B_1} -b_1\right).
\ee
Equating $\ell_2 = \ell_3 = \pi \phi_b/\phi_r$, we find
\be
B_1 \approx \sqrt{\frac{m\phi_r}{2+\pi}},\qquad M_w \approx 2\frac{B_1}{b_1}\approx 2\phi_r.
\ee
The observer mass on the wormhole side is significantly larger than $M$, and significantly back-reacts on $\phi$. 

Finally, we can also put constraints on the time between various events along the worldline. For instance, the time between an insertion of $\chi$ and the reheating. This allows generalizing to asymmetric insertions of $\chi$ along the worldline, relevant for $|G_\chi(t-i\ep)|^2$. In this case the observer mass will have different asymptotic values.
\section{Summary and speculations}
We have studied an apparently well formulated question, namely the correlation function $G_\chi(\tau)$ measured by a dynamical observer in de Sitter, equipped, \`a la \cite{Chandra,Witten}, with a clock. Semi-classical computation of this observable gives an exponentially decaying answer $\propto e^{-t/2}$, where $t={\rm Re}(\tau)$. When instead computing $|G_\chi(\tau)|^2$ in JT gravity, we found, in addition to the square of the decaying answer $\propto e^{-t}$, a new {\em non-decaying} saddle of the gravitational path integral that is suppressed by $e^{-S_{\rm dS}}$. This saddle consists of a geometry that does not factorize into separate computations of $G_\chi(\tau)$ and $G_\chi(\tau)^*$, but it connects the $\ket$ in one to the $\bra$ in the other. It can be visualized as two Lorentzian cones connected by two Lorentzian ``double tops'', regularized near the tips by going in the complex plane as in \eqref{rho_contour}. It is somewhat reminiscent of the double cone geometry considered in \cite{SSS1}.

Currently, there is no clear-cut rule for deciding which saddles of the gravitational path integral are to be summed over. One possible criterion is to only allow metrics on which $p$-form quantum field theories are well-defined \cite{Witten_KS}. This has been shown to imply the following constraint on the eigenvalues $\{\lambda_i\}$ of the metric \cite{Kontsevich}
\be
\sum_i |{\rm Arg}(\lambda_i)| <\pi,
\ee
after choosing $-\pi< {\rm Arg}(\lambda_i)\leq \pi$ for each eigenvalue. This is not satisfied by the new saddle.\footnote{We thank Yiming Chen for pointing this out.} Regardless of how the $t$ contour in \eqref{rho_contour} is deformed it has to cross the line ${\rm Im}(t)=i\pi/2$, at which ${\rm Arg}(\sinh^2 t) = \pi$.

If this saddle is nevertheless allowed, given that it does not contribute to $G_\chi(\tau)$, it can be interpreted as signaling a transition to a chaotic phase where the correlation function fluctuates around zero. This is a well-known phenomenon in theories with a finite dimensional Hilbert space. However, the time-scale we are getting for the transition $t\sim S_{\rm dS}$ is of the order of the ``Page time'' suggested in \cite{Arkani}. It is much shorter than the typical expectation $e^{S_{\rm dS}}$, of the same order as the macroscopic Poincar\'e recurrence time.\footnote{Curiously, a non-geometric example (Brownian SYK) exhibits a rapid transition into the chaotic phase at $t\sim S$ \cite{SSS1}.} 

This is apparently at odds with the late time behavior of correlation functions on AdS black hole background, where one can reliably compute this time in JT gravity \cite{Saad} and get the expected answer. Indeed, the correlation function decays exponentially at early times, but after $t\sim \beta$, gravitational interactions on the disk topology become strongly coupled and change the exponential decay of $G_\chi(\tau)$ into a power-law. It is only after $t\sim e^{S_{\rm BH}/2}$ that the topologically nontrivial handle-disk geometries (related to the double cone of \cite{SSS1}) stop the decay and lead to a ramp in $G_\chi(\tau)$. 

In higher dimensions, it was originally observed that thermal AdS saddle can change the decaying behavior of the correlation function after $t\sim S_{\rm BH}$ \cite{Maldacena_eternal}. Additional gravitational saddles were also found in \cite{Fitzpatrick} for BTZ black holes. These are both examples of a change that is visible in $G_\chi(\tau)$ after $t\sim S_{\rm BH}$. The need for an additional contribution to account for the chaotic behavior was emphasized early on in \cite{Barbon}. 

Could gravitational fluctuations in dS similarly modify the decay of $G_\chi(\tau)$? We cannot rule out this possibility, but at least in the JT example, this has to happen in a different way compared to AdS. While in the AdS computation large Lorentzian time pushes the matter correlation function on the disk topology into the regime of large back-reaction, on the dS side it corresponds to sending the UV cutoff to infinity, under which the answer for every topology reaches a well-defined limit. 


It would be interesting to check if higher genus contributions to $G_\chi(\tau)$ can change its time-dependence. Perhaps this could happen if such geometries contain shortcuts through which the two $\chi$ insertions can communicate, as in the AdS case \cite{Saad}. Alternatively, the dominance of the $|G_\chi(\tau)|^2$ wormhole at $t\sim S_{\rm dS}$ might be pointing to a special feature of the holographic dual. 

From a holographic perspective, the transition of AdS correlators from exponential to power-law decay is due to the sharp edge of the density of states $\propto \sqrt{E}$, and the ramp is due to a particular spectral correlation, namely the repulsion of energy levels \cite{SSS1}. The double cone geometry that leads to a ramp in the bulk computation is {\em not} a saddle point of the gravitational path integral. On the other hand, an exponentially suppressed plateau contribution to the ensemble average of $|G_\chi(\tau)|^2\propto e^{-2 S}$ can be argued for based on eigenstate thermalization hypothesis but without invoking any spectral correlations. In \cite{Stanford}, this has been shown to arise in AdS$_2$ from wormhole saddles that are stabilized by the matter fields. 

Pure analogy would suggest that the decay rate of $G_\chi(\tau)$ could remain exponential in dS because of the absence of a sharp edge in the spectrum, in which case the non-perturbative contributions become visible after $t\sim S_{\rm dS}$. A ramp, even if present on the dS side, would have been missed in our search for stabilized wormholes. But what would be the reason for the much higher plateau we are finding $|G_\chi(\tau)|^2\propto e^{- S_{\rm dS}}$ instead of $\propto e^{-2S_{\rm dS}}$?

Of course, it is perfectly possible that there is no holographic dual, or there is one, but the semi-classical computation simply fails at $t\sim S_{\rm dS}$ without giving any hint at the nature of such a dual. 

\vspace{0.3cm}
\noindent
\section*{Acknowledgments}

I am grateful to Victor Gorbenko, Shota Komatsu, Kyriakos Papadodimas, Eva Silverstein, Giovanni Villadoro and especially to Lorenzo Di Pietro for enlightening me at various phases of this project. I am also grateful to the participants and organizers of the CERN workshop on Cosmology, Quantum Gravity, and Holography, during which this project was initiated. I thank Yiming Chen and Oliver Janssen for pointing out several errors in the earlier versions.
\appendix 
\section{dS correlators and geodesics}
In two dimensions the correlation function of a scalar field in the Hartle-Hawking state is given by 
\be\label{2pf2}
\expect{\chi(X)\chi(Y)} = \frac{\Gamma(\frac{1}{2}+ i\nu) \Gamma(\frac{1}{2}- i \nu)}{4\pi} \hyp(\frac{1}{2}+ i\nu,\frac{1}{2}-i\nu, 1 , z),
\ee
where $\nu \equiv \sqrt{m^2 - \frac{1}{4}}$, and using the embedding of dS$_2$ in $3d$ Minkowski with the condition $X\cdot X = Y\cdot Y = -1$
\be
z  = \frac{1+ X \cdot Y}{2}.
\ee
Any distinct pair of points $X$ and $Y$ in dS can be connected either by a spacelike geodesic of length $s\in (0,\pi)$, in which case $0\leq z\leq 1$, or a timelike geodesic of length $s = it$, equivalent to $z>1$, or a complex geodesic of length $s = \pi + i t$, equivalent to $z<0$. The latter case occurs when $X$ is spacelike and superhorizon with respect to $Y$. The complex geodesic consists of a timelike piece that extends from $Y$ to the $t=0$ slice in the metric \eqref{dS2}, Wick rotates into a spacelike geodesic on the 2-sphere \eqref{S2}, and after moving a distance $\pi$ Wick rotates back and moves forward in time to reach $X$. By dS isometries such complex geodesics can all be mapped to the one shown in figure \ref{fig:hh}. Moreover, when computing in-in correlation functions, there are analogous geodesics that connect $X$ and $Y$ by extending in the $\bra$ side. 

We now use them to write approximate formulas for $G_\chi$, away from the boundaries of the above three configurations. In the Euclidean regime
\be\label{GEuc}
G_\chi(z = \cos^2(s/2)) \approx \frac{1}{2\sqrt{2\pi \nu \sin s} }\left[e^{-\nu s}- e^{-\nu (2\pi - s)}\right],
\ee
in the Lorentzian region with $t\gg 1$
\be
G_\chi(z = \cosh^2(t/2)) \approx -\frac{i e^{-t/2}}{2\sqrt{\pi \nu}} \left[e^{-i\nu t + i\frac{\pi}{4}}+ e^{-2\pi \nu+ i\nu t -i \frac{\pi}{4}}\right]
\ee
and in the superhorizon region
\be
G_\chi(z = -\sinh^2(t/2)) \approx \frac{ e^{-t/2-\pi \nu}}{\sqrt{\pi \nu}}\cos\left(\nu t -\frac{\pi}{4}\right).
\ee
Up to an overall constant normalization, and a relative phase that is fixed by the appropriate matching condition at the turning points, these results coincide with the sum of the saddle points of the point particle action. There are always two saddles. In the Euclidean regime they correspond to the two geodesics that connect any two (non-antipodal) points on the sphere. To study fluctuations around the saddle, we parameterize the sphere as
\be
ds^2 = (1-x^2) d\tau^2 + \frac{dx^2}{1-x^2},
\ee
where $x=0$ corresponds to the classical solution. The action is given by
\be
S_{\rm pp} = - m\int d\tau \sqrt{(1- x^2) + \frac{\dot x^2}{1-x^2}}.
\ee
The classical action is $-m \tau$, and the 1-loop determinant $\propto 1/\sqrt{\sin \tau}$ (see e.g. Appendix 1 in section 7 of \cite{Coleman_aspects}). Note that in the large mass limit $\nu = m - \frac{1}{8 m} + \cdots$. The difference between the classical actions and the exponents in \eqref{GEuc} is of the order of the higher curvature corrections that can be added on the worldline $\Delta S_{\rm pp} = \frac{1}{m} \int ds R$.

\bibliography{bibwl}
\end{document}